\documentclass[prd, preprint, aps]{revtex4}
\usepackage{latexsym, graphicx}
\usepackage{amsfonts,amsmath,amssymb,bm}

\def\be{\begin{equation}}
\def\ee{\end{equation}}

\def\te{\end{equation}}
\def\bea{\begin{eqnarray}}
\def\nn{\nonumber\\}
\def\tea{\end{eqnarray}}
\begin{document}

\title{Linking the hydrodynamic and kinetic description of a dissipative relativistic conformal theory}

\author{E. Calzetta}
\email{calzetta@df.uba.ar}
\affiliation{CONICET and Departamento de F\'isica, Facultad de Ciencias Exactas y Naturales, Universidad de Buenos Aires-Ciudad Universitaria, Pabell\'on I, 1428 Buenos Aires, Argentina}

\author{J. Peralta-Ramos}
\email{jperalta@ift.unesp.br}

\affiliation{Instituto de F\'isica Te\'orica, Universidade Estadual Paulista, 
Rua Dr. Bento Teobaldo Ferraz 271 - Bloco II, 01140-070 S\~ao Paulo, Brazil}

\date{\today}

\begin{abstract}
We use the entropy production variational method to associate a one particle distribution function to the assumed known energy-momentum and entropy currents  describing a relativistic conformal fluid. Assuming a simple form for the collision operator we find this one particle distribution function explicitly, and show that this method of linking the hydro and kinetic description is a non trivial generalization of Grad's ansatz. The resulting constitutive relations are the same as in the conformal dissipative type theories discussed in J. Peralta-Ramos and E. Calzetta, Phys. Rev. D {\bfseries 80}, 126002 (2009). 
Our results may prove useful in the description of freeze-out in ultrarelativistic heavy-ion collisions.

\end{abstract}

\maketitle

\section{Introduction}
The relationship between the kinetic and hydrodynamic descriptions of a dissipative relativistic system is a long standing puzzle, because while the relativistic Boltzmann equation is well understood \cite{libro,is72}, its obvious match, namely the Eckart and Landau-Lifshitz theories \cite{first}, are plagued with causality and stability problems. Solving this puzzle has acquired a certain urgency, because being able to associate a one particle distribution function to known hydrodynamic currents is an essential step in describing freeze-out in hydrodynamic models of heavy ion collisions (see for instance \cite{rom09,luz,dus,den,mon}).

There is a long history in the development of different formalisms to derive hydrodynamics from kinetic theory, the most well-known methods being Grad's method of moments and the Chapmann-Enskog expansion \cite{groot,stew,libro,gorban,israel,is72,grad,chap,cerc}. For some recent theoretical developments see \cite{den,mon,ktoh}.

Here, we will use the entropy production variational methods (EPVM) to provide a linkage between the kinetic and hydrodynamic description; for a review see \cite{mep}. The idea of entropy production variational methods is to find the particle distribution function which extremizes the entropy production, providing in the process a means of closing the infinite chain of hydrodynamic equations (i.e. providing a {\it closure}) \cite{mep,chris}. 

In order to carry out this derivation explicitly we assume, for simplicity, a conformally invariant theory and use a linear collision operator. This operator satisfies energy-momentum conservation and guarantees the $H$ theorem, ensuring that the resulting theory exactly satisfies the Second Law. The size of the collision term is determined by a dimensionful parameter $\tau$ with the physical meaning of a relaxation time. There is a natural expansion of the solution in powers of $\tau$. We shall find explicitly the one-particle distribution function singled out by the EPVM to second order in $\tau$.

The most important result from this paper is the comparison of the distribution function picked up by the EPVM and the better known solution provided by Grad's ansatz. We shall see that they differ in two main ways. First, the EPVM is not tied up to a gradient expansion in the hydrodynamical variables, and so it becomes an attractive choice in situations where those gradients are expected to be large. Second, the Grad ansatz is usually applied in conjunction with simplifying assumptions regarding the Boltzmann equation, to the effect that the resulting correction to the distribution function is expressed solely in terms of the shear viscosity and the equilibrium energy density and pressure; it bears no memory of the collision term, unless through a global factor. The correction to the distribution function derived from the EPVM, on the other hand, depends in an essential way on the collision term. Indeed this result could be seen as a way to associate a simple kinetic equation to nontrivial freeze-out prescriptions derived from more fundamental physics and discussed in \cite{dus} (see also \cite{luz}).

The EPVM provides a prescription to associate a one-particle distribution function to given macroscopic currents, but gives no information on those currents or their further time evolution. To fill this gap, we show that the EPVM may be cast in the framework of conformal divergence-type theorys (DTTs). These theories were discussed in \cite{nos,nosprc}. In this way we determine the dynamics of the energy-momentum and entropy currents; the former is conserved, while entropy production is equated to its extremal value. The association with DTTs, moreover, affords a simple way of investigating the causality and thermodynamic consistency of the hydrodynamics associated to the EPVM.

As a byproduct, the derivation given here provides a novel kinetic interpretation of the nonequilibrium tensor of DTTs, as a Lagrange multiplier enforcing the energy-momentum constraint when extremizing the entropy production. 

Divergence-type theories \cite{geroch} are hydrodynamic theories which are based on extending the set of hydrodynamic variables used to describe a nonequilibrium system to include a traceless tensor which vanish in equilibrium. These are {\it exact} hydrodynamic theories in the sense that they are not based on gradient expansions, and therefore can describe situations with large gradients (i.e. shock-waves) in which the so-called second-order theories (SOTs) \cite{jou,israel,brss} are known to fail. In \cite{nos} we have developed a quadratic DTT for a conformal field, which reproduces the SOT developed in \cite{brss} when velocity gradients are small. We then applied it in \cite{nosprc} to describe the evolution of the fireball created in ultrarelativistic heavy-ion collisions.

The paper is organized as follows. In section \ref{var} we develop the EPVM approach. We obtain the variational equation for the entropy production, and then we describe the linear collision term used here and solve the variational equation perturbately up to second order in the relaxation time. In section \ref{ansatz} we compare those results to Grad's quadratic ansatz. In section \ref{dtt} we overview DTTs and give a brief summary of the main results for conformal fields, and make explicit the connection between the EPVM and DTTs. Finally, we summarize our results in section \ref{summ}.

\section{Entropy production: variational method}
\label{var}

In this section we set up the problem within the context of kinetic theory and derive the variational equation to be solved perturbately. We then specify the collision operator and solve the variational problem to second order in the relaxation time.

\subsection{The variational equations}

We consider a relativistic kinetic theory in flat space-time \cite{groot,stew,libro,israel,is72}. We use signature $\left(-,+++\right)$. The distribution function $f(x^\mu,p^\mu)$ determines the energy momentum tensor $T^{\mu\nu}$ and the entropy current $S^{\mu}$

\be
T^{\mu\nu}=\int\:Dp\:p^{\mu}p^{\nu}f
\te
and 
\be
S^{\mu}=\int\:Dp\:p^{\mu}\left\{\left(1+f\right)\ln\left[1+f\right]-f\ln\left[f\right]\right\}
\te

with
\be
Dp=\frac{d^4p}{\left(2\pi\right)^4}\theta\left(p^0\right)\rho\left(-p^2\right)
\te
where $\rho$ is the density of states and $-p^2=\left(p^0\right)^2-\mathbf{p}^2$. We assume a conformal theory where $\rho\left(-p^2\right) \propto \delta\left(-p^2\right)$. The distribution function obeys a Boltzmann-like equation, which we write in compact form as

\be
p^{\mu}f_{,\mu}=I_{col}\left[f\right] ~.
\te
Energy-momentum conservation implies the constraint

\be
\int\:Dp\:p^{\mu}I_{col}\left[f\right]\equiv 0
\te
identically in $\mu$ and in $f$, while entropy creation reads

\be
S^{\mu}_{,\mu}=\int\:Dp\:I_{col}\left[f\right]\ln\left[1+f^{-1}\right] ~.
\te

Given a vector $\beta_{\mu}$ we define the thermal distribution

\be
f_0=\frac1{e^{-\beta p}-1}
\te
with $\beta p=\beta_{\mu}p^{\mu}$. It is convenient to parameterize deviations from thermal equilibrium as follows 

\be
f=f_0\left[1+\left(1+f_0\right)\chi\right] ~.
\label{chidef}
\te
Then 

\be
T^{\mu\nu}=T_0^{\mu\nu}+\Pi^{\mu\nu}
\label{temunu}
\te
where $T_0^{\mu\nu}$ corresponds to a perfect fluid, and

\be
\Pi^{\mu\nu}=\int\:D_{\beta}p\:p^{\mu}p^{\nu}\chi
\te
with 
\be
D_{\beta}p=Dp\:f_0\left(1+f_0\right) ~.
\te
It is convenient to introduce the notation
\begin{equation}
\left\langle \cdots \right\rangle =  \int\:D_{\beta}p\:(\cdots) ~.
\label{braket}
\end{equation}
In this notation $\Pi^{\mu\nu}=\left\langle p^{\mu}p^{\nu}\chi \right\rangle$

We have

\be
S^{\mu}=S_0^{\mu}-\beta_{\nu}\Pi^{\mu\nu}+S_1^{\mu}=p\beta^{\mu}-\beta_{\nu}T^{\mu\nu}+S_1^{\mu}
\label{entroflux}
\te
where $p=\rho /3$ is the pressure and
\be
S_1^{\mu}=\int\:D_{\beta}p\:p^{\mu}\left\{\left(f_0^{-1}+\chi\right)\ln\left[1+f_0\chi\right]-\left(\left(1+f_0\right)^{-1}+\chi\right)\ln\left[1+\left(1+f_0\right)\chi\right]\right\} ~.
\te
Similarly

\be
S^{\mu}_{,\mu}=\int\:Dp\:I_{col}\left[f\right]\left\{\ln\left[1+f_0^{-1}\right]+\ln\left[\frac{1+f_0\chi}{1+\left(1+f_0\right)\chi}\right]\right\} ~.
\te
But the first term integrates identically to zero, so

\be
S^{\mu}_{,\mu}=\int\:D_{\beta}p\:I_{\beta}\left[\chi\right]\ln\left[\frac{1+f_0\chi}{1+\left(1+f_0\right)\chi}\right]
\te
where

\be
I_{\beta}\left[\chi\right]=\frac{I_{col}\left[f_0\left[1+\left(1+f_0\right)\chi\right]\right]}{f_0\left(1+f_0\right)} ~.
\te

By the way, we notice two identities. By taking a variation of the energy-momentum conservation constraint, we get

\be
\int\:D_{\beta}q\:q^{\mu}\:\frac{\delta I_{\beta}\left[\chi\right]\left(q\right)}{\delta\chi\left[p\right]}=0 ~.
\te
On the other hand, observe that an infinitesimal $\delta\chi=\delta\beta_{\mu}p^{\mu}$ is just a shift in $\beta$, and therefore  $I_{\beta}\left[\delta\chi\right]\left(p\right)=0$ identically in $p$. Expanding to first order in $\delta\beta$ we get

\be
\int\:D_{\beta}q\:\frac{\delta I_{\beta}\left[\chi\right]\left(p\right)}{\delta\chi\left[q\right]}\:q^{\mu}=0  ~.
\te
In other words, $p^{\mu}$ is both a right and left null eigenvector of $\delta I_{\beta}\left[\chi\right]\left(q\right)/{\delta\chi\left[p\right]}$.

Below we shall restrict ourselves to the case of linear collision terms, these being essentially the only ones for which a closed-form solution may be found. The form of the entropy production suggests that to conform to the $H$ theorem, such a functional must be linear not in $\chi$ but in the new variable

\be
\zeta=\ln\left[\frac{1+\left(1+f_0\right)\chi}{1+f_0\chi}\right] ~.
\label{zetanew}
\te
This relation may be inverted to yield

\be
\frac{1+\left(1+f_0\right)\chi}{1+f_0\chi}=e^{\zeta}
\te
so 
\be
\chi=\frac{e^{\zeta}-1}{1-f_0\left(e^{\zeta}-1\right)}
\te
or else, expanding to second order in $\zeta$

\be
\chi =\zeta+\frac12\left(1+2f_0\right)\zeta^2 ~.
\te

Suppose now we wish to find the distribution function that extremizes $S^{\mu}_{,\mu}$ given the values $\bar{T}^{\mu\nu}$ and $\bar{S}^{\mu}$ of the energy momentum tensor and entropy current. Choosing a suitable $\beta$ and decomposing both the kinetic theory and the hydrodynamic currents as above, we end up solving the variational problem

\be
\frac{\delta }{\delta\zeta\left[p\right]}\left[S^{\mu}_{,\mu}-\lambda_{\mu}S_1^{\mu}-\lambda_{\mu\nu}\Pi^{\mu\nu}\right]=0
\label{vareq}
\te
where $\lambda_{\mu}$ and $\lambda_{\mu\nu}$ are Lagrange multipliers enforcing the constraints. Observe that $\lambda_{\mu\nu}$ is dimensionless while $\lambda_{\mu}$ has dimensions of temperature. Below we shall restrict ourselves to the case $\lambda_{\mu}=0$.

\subsection{The collision integral}
\label{coll}

It is clear that to actually solve for $\zeta$ we need to know something about the collision operator. In this section we shall investigate the structure of linear operators \cite{groot,stew,libro}. In principle one would like to write  $I_{\beta}\left[\zeta\right]=-F\zeta\left(p\right)/\tau$, where $\tau$ is the relaxation time. Observe that $F$ has dimensions of temperature, and $\tau$ has dimensions of time, namely inverse temperature. To avoid picking up a preferred direction in the rest frame, it is natural to request that $F=F\left[\omega_p\right]$, where $\omega_p=-u_{\mu}p^{\mu}$. However, these restrictions are not sufficient, since such a kinetic equation would violate the energy-momentum conservation constraints. 

To preserve the momentum constraints we introduce a projection operator $Q$ such that
\be
\left\langle p^{\mu}Q\left[f\right]\right\rangle=0
\te
for any $f$, but $Q\left[f\right]=f$ if $\left\langle p^{\mu}f\right\rangle=0$. The notation $\left\langle \right\rangle$ is defined in eq. (\ref{braket}).
In the rest frame, we write

\be
Q\left[f\right]=f-\frac1{\left\langle \omega_q^2\right\rangle} \left[\omega_p\left\langle \omega_qf\right\rangle+ 3p^i\left\langle q_if\right\rangle\right] 
\te
where we exploit the fact that for a conformal theory $\left\langle p^�p^j\right\rangle =\delta^{ij}\left\langle \omega_p^2\right\rangle /3$.  Now we write the collision integral as 

\be
I_{\beta}\left[\chi\right]=\frac{-1}{2\tau}Q\left[FQ\left[\zeta\right]\right]
\te
Suppose we wish to solve the equation

\be
Q\left[FQ\left[g\right]\right]=f
\label{pattern}
\te
where $\left\langle p^{\mu}f\right\rangle=0$. Then, in the rest frame

\be
FQ\left[g\right]=f-A\omega_p-B_ip^i  ~.
\te
The constants $A$ and $B_i$ must enforce the integrability conditions

\be
\left\langle \frac{\omega_p}{F}\left(f-A\omega_p-B_ip^i\right)\right\rangle=\left\langle \frac{p^j}{F}\left(f-A\omega_p-B_ip^i\right)\right\rangle=0 ~.
\te
Therefore

\be
A=\frac{\left\langle \omega_pf/F\right\rangle}{\left\langle \omega_p^2/F\right\rangle} ~,
\te

\be
B^j=3\frac{\left\langle p^jf/F\right\rangle}{\left\langle \omega_p^2/F\right\rangle}
\te
and

\be
g=\frac1F\left(f-A\omega_p-B_ip^i\right) ~.
\te
We will make use of this properties in what follows. 

The linear collision operator used here is quite general, and it is interesting to note that for $F= T$ we recover Marle's relativistic generalization of the BGK model \cite{marle}, while for $F=\omega_p$ we get the Anderson-Witting model \cite{ander} (see \cite{taka} for a comparison of these kinetic models to Israel-Stewart formalism). 

\subsection{Perturbative solution}
\label{pert}

The structure of the collision term suggests we seek a solution as an expansion in powers of the relaxation time $\tau$. We shall consider the solution up to second order.

The equation we wish to solve is

\be
Q\left[FQ\left[\zeta\right]\right]=\tau\lambda_{\mu\nu}p^{\mu}p^{\nu}\left[1+\left(1+2f_0\right)\zeta\right] ~.
\label{eqlin}
\te
We expand 

\be
\zeta=\zeta_1+\zeta_2
\te
and 
\be
\lambda_{\mu\nu}=\lambda^{(0)}_{\mu\nu}+\lambda^{(1)}_{\mu\nu} ~.
\label{expandepvm}
\te
Then we find the equations

\be
Q\left[FQ\left[\zeta_1\right]\right]=\tau\lambda^{(0)}_{\mu\nu} p^{\mu}p^{\nu}
\te
and 
\be
Q\left[FQ\left[\zeta_2\right]\right]=\tau p^{\mu}p^{\nu}\left[\lambda^{(1)}_{\mu\nu}+\left(1+2f_0\right)\lambda^{(0)}_{\mu\nu}\zeta_1\right]
\te
and the integrability conditions

\be
\lambda^{(0)}_{\mu\nu}\left\langle p^{\mu}p^{\nu}p^{\rho}\right\rangle=0
\te
and 
\be
\lambda^{(0)}_{\mu\nu}\left\langle \left(1+2f_0\right)\zeta_1\left(p\right)p^{\mu}p^{\nu}p^{\rho}\right\rangle
+\lambda^{(1)}_{\mu\nu}\left\langle p^{\mu}p^{\nu}p^{\rho}\right\rangle=0 ~.
\te
The correction to the one particle distibution function reads

\be
\chi=\zeta_1+\zeta_2+\frac12\left(1+2f_0\right)\zeta_1^2 ~.
\label{chi12}
\te
Therefore for the energy momentum tensor we shall find

\be
\Pi^{\mu\nu}=\Pi_1^{\mu\nu}+\Pi_2^{\mu\nu}
\te
where 
\be
\Pi_1^{\mu\nu}=\int\:D_{\beta}p\:p^{\mu}p^{\nu}\zeta_1
\te
and 
\be
\Pi_2^{\mu\nu}=\int\:D_{\beta}p\:p^{\mu}p^{\nu}\left[\zeta_2+\frac12\left(1+2f_0\right)\zeta_1^2\right] ~.
\te
The entropy flux

\be
S_1^{\mu}=\frac{-1}2\int\:D_{\beta}p\:p^{\mu}\chi^2=\frac{-1}2\int\:D_{\beta}p\:p^{\mu}\zeta_1^2
\te
so there is no first order correction.  
We shall only consider the first nonvanishing contribution to the entropy creation

\be
S^{\mu}_{,\mu}=\lambda^{(0)}_{\mu\nu}\int\:D_{\beta}p\: p^{\mu}p^{\nu}\zeta_1
\te
which is already quadratic in deviations from equilibrium.

We will now go over to calculate $\zeta_1$ and $\zeta_2$. 

\subsubsection{First order solution}
Let us now consider the first order solution for the specific collision term introduced above. To make things simpler, we shall go to the rest frame, where the integrability conditions become

\be
\lambda^{(0)}_{00}+\frac13\lambda^{(0)i}_i=0
\te
and 
\be
\lambda^{(0)}_{0i}=0  ~.
\te
For simplicity and without loss of generality we shall assume that $\lambda^{(0)}_{00}=\lambda^{(0)i}_{i}=\lambda^{(0)}_{0i}=0$. 

The solution reads

\be
\zeta_1=\frac{\tau}F\lambda^{(0)}_{ij}p^{i}p^{j} ~.
\label{zeta1sol}
\te
In our particular case, $A=B_i=0$.

If we use this to compute $\Pi_1^{\mu\nu}$ we get $\Pi_1^{00}=\Pi_1^{0i}=0$. To compute $\Pi_1^{ij}$ recall that for any $G\left[\omega\right]$

\be
\left\langle G\left[\omega_p\right]p^{i}p^{j}p^{k}p^{l}\right\rangle=\frac{\left\langle G\left[\omega_p\right]\omega_p^4\right\rangle}{15}\left[\delta^{ij}\delta^{kl}+\delta^{ik}\delta^{lj}+\delta^{il}\delta^{jk}\right]
\te
so

\be
\Pi_1^{ij}=\frac{2\tau}{15}\left\langle \frac{\omega_p^4}F\right\rangle\lambda^{(0)ij}
\label{pi1mepp}
\te
which is traceless. 

The lowest order nontrivial contribution to the entropy flux is 

\be
S_1^0=\frac{-\tau^2}{15}\left\langle \frac{\omega_p^5}{F^2}\right\rangle\lambda^{(0)}_{ij}\lambda^{(0)ij}
\label{entroepvm}
\te
and $S_1^i=0$ in the rest frame.
Similarly

\be
S^{\mu}_{,\mu}=\frac{2\tau}{15}\left\langle \frac{\omega_p^4}F\right\rangle\lambda^{(0)}_{ij}\lambda^{(0)ij} ~.
\label{divs1}
\te

\subsubsection{Second order solution}
\label{sos}

We now use the first order solution to investigate the second order one. Let us start by writing the consistency condition in the rest frame

\be
\tau^{-1}\left\langle \left(1+2f_0\right)F\zeta_1^2p^{\rho}\right\rangle
+\lambda^{(1)}_{\mu\nu}\left\langle p^{\mu}p^{\nu}p^{\rho}\right\rangle=0 ~.
\te
If $\rho=0$, then

\be
\left\langle \omega_p^3\right\rangle\left[\lambda^{(1)}_{00}+\frac13\lambda^{(1)k}_{k}\right]=-\tau^{-1}\left\langle \left(1+2f_0\right)F\zeta_1^2\omega_p\right\rangle\neq 0
\te
and if $\rho=i$

\be
\left\langle \omega_p^3\right\rangle\lambda^{(1)}_{i0}=0 ~.
\te
We see that $\lambda^{(1)}_{\mu\nu}$ may be transverse but not both transverse and traceless.

The point is that any term in $\lambda^{(1)}_{\mu\nu}$ which is not strictly required by the consistency conditions may be absorbed into $\lambda^{(0)}_{\mu\nu}$, and so there is no loss of generality if we simply take

\be
\lambda^{(1)ij}=-\tau^{-1}\Lambda\delta^{ij}
\label{l1ij}
\te
with 
\be
\Lambda = \frac{\left\langle \left(1+2f_0\right)F\zeta_1^2\omega_p\right\rangle}{\left\langle \omega_p^3\right\rangle} ~.
\te
The second order equation then reads

\be
Q\left[FQ\left[\zeta_2\right]\right]=- \omega_p^2\Lambda+\left(1+2f_0\right)F\zeta_1^2 ~.
\te
Observe that the right hand side vanishes if integrated against $\omega_p$, so the equation may be solved, but not if integrated against $\omega_p/F$. Thus the solution is

\be
\zeta_2=\left(1+2f_0\right)\zeta_1^2- \Lambda\frac{\omega_p^2}F-A\frac{\omega_p}F
\label{zeta2sol}
\te
where

\be
A={\left\langle \frac{\omega_p^2}F\right\rangle}^{-1}{\left[\left\langle \left(1+2f_0\right)\omega_p\zeta_1^2\right\rangle- \Lambda\left\langle \frac{\omega_p^3}F\right\rangle\right]} ~.
\te

So far, we have

\be
\Pi_2^{\mu\nu}=\int\:D_{\beta}p\:p^{\mu}p^{\nu}\left[\frac32\left(1+2f_0\right)\zeta_1^2- \Lambda\frac{\omega_p^2}F-A\frac{\omega_p}F\right]
\te
which is traceless. However, this cannot be the true correction to the energy-momentum tensor because $\Pi_2^{00}\neq 0$. This means that the parameter $T$ in our equations is not the physical temperature $T_{phys}$, but rather

\be
T=T_{phys}-\delta T
\label{tshift}
\te
where $\delta T$ is chosen so that

\be
\frac{d\rho}{dT}\delta T=4\rho\frac{\delta T}T=\Pi_2^{00} ~.
\te
The physical correction to the energy-momentum tensor $\Pi_{2,phys}^{00}=\Pi_{2,phys}^{0i}=0$ and

\be
\Pi_{2,phys}^{ij}=\Pi_2^{ij}-\frac13\delta^{ij}\Pi_2^{00}
\te
which is still traceless. 

Now there is only one traceless tensor quadratic in $\lambda^{(0)}_{ij}$, and so we must have

\be
\Pi_{2,phys}^{ij}=K\left\{\lambda_{m}^{(0)i}\lambda^{(0)mj}-\frac13\delta^{ij}\lambda^{(0)}_{mn}\lambda^{(0)mn}\right\}
\label{pi2mepp}
\te
where

\be
K=12\tau^2\left\langle \left(1+2f_0\right)\frac{\omega_p^6}{F^2}\right\rangle
\label{kconst}
\te
It is interesting to recall the identities

\be
\left\langle \left(1+2f_0\right)\omega_p^6\right\rangle=T^2\frac d{dT}\left\langle \omega_p^5\right\rangle=7T\left\langle \omega_p^5\right\rangle
\te

\be
\left\langle \left(1+2f_0\right)\omega_p^4\right\rangle=T^2\frac d{dT}\left\langle \omega_p^3\right\rangle=5T\left\langle \omega_p^3\right\rangle=5T^3\frac d{dT}\rho=15T^2\left(\rho +p\right)
\te
Therefore for the Marle collision term $F=T$ we get

\be
K_{Marle}=84\tau^2T^{-1}\left\langle \omega_p^5\right\rangle
\label{kmarle}
\te
and for the Anderson and Witting collision term $F=\omega_p$ \cite{ander,taka}

\be
K_{AW}=180\tau^2T^2\left(\rho +p\right)
\label{kaw}
\te
In section \ref{dtt} we will see that this is precisely the form of $\Pi_2^{\mu\nu}$ obtained in the quadratic DTT.

For clarity, we will briefly summarize the main logical steps followed in this section. For a linear collision term the variational equation (\ref{vareq}) becomes (\ref{eqlin}). If we expand the Lagrange multiplier $\lambda_{\mu\nu}$ and the nonequilibrium correction to the distribution function (now parametrized by the new variable $\zeta$ given by (\ref{zetanew}) to satisfy the $H$ theorem), we can solve the variational equation perturbately up to second order in the relaxation time. The first and second order solutions are given by (\ref{zeta1sol}) and (\ref{zeta2sol}), respectively. We emphasize that the assuptions made regarding the integrability conditions, namely  $\lambda^{(0)}_{00}=\lambda^{(0)i}_{i}=\lambda^{(0)}_{0i}=0$ and that leading to (\ref{l1ij}), constitute no loss of generality in the development.

\section{Comparison to Grad's ansatz}
\label{ansatz}
We will now compare, in the context of the method of moments, the closure provided by the entropy production variational equation with that provided by Grad's quadratic ansatz \cite{grad} (see also \cite{stew,groot}; for a recent generalization to multicomponent systems see \cite{mon}). 

Taking moments of the Boltzmann equation one obtains an infinite set of equations:
\be
\begin{split}
\partial_\mu \int\:Dp\:p^{\mu} f &\equiv N^\mu_{,\mu} = \int\:Dp\: I_{col}[f] = 0 \\
\partial_\mu \int\:Dp\:p^{\mu}\:p^\nu f &\equiv T^{\mu\nu}_{,\mu} = \int\:Dp\:p^\nu I_{col}[f] = 0 \\
\partial_\mu \int\:Dp\:p^{\mu}\:p^\nu\:p^\delta f &= \int\:Dp\:p^\nu \:p^\delta I_{col}[f] \\
\ldots
\end{split}
\label{moments}
\te
where in the first and second lines we have used that $(1,p^\mu)$ are collisional invariants. This infinite set is completely equivalent to Boltzmann equation. 

The method of moments rests on the assumption that a finite subset of this hierarchy will give a reasonable description of the hydrodynamic regime. The most common case is to consider only the first three equations of the hierarchy. However, the truncated system is not a closed one, since the derivative of the second moment can not be expressed solely in terms of the hydrodynamic variables $(N^\mu,T^{\mu\nu})$. In order to close the system, and following Grad's idea, Israel and Stewart \cite{israel} proposed expanding the single-particle distribution $f=f_0+\delta f$ around its equilibrium value in a Taylor-like  series in $p^\mu$ and truncating it at quadratic order. In the notation of eq. (\ref{chidef}) above, this is
\be
\chi= C+C^\mu p_\mu + C^{\mu\nu} p_\mu p_\nu+\ldots
\label{exp}
\te
where the coefficients $(C,C^\mu)$ correspond to a shift in chemical potential, temperature and velocity, and may be taken as zero. The coefficient $C^{\mu\nu} $ is subject to the constraint that we must reproduce eq. (\ref{temunu}). It therefore has to be traceless and transverse. In the rest frame, the nontrivial components satisfy

\be 
2\left\langle C^{ij}\omega_p^{4}\right\rangle=\Pi^{ij} ~.
\label{constraint}
\te
This constraint does not determine $C^{ij}$ by itself, and so we must resort to eq. (\ref{moments}) or else to attempt a solution to the Boltzmann equation.
The first two equations in (\ref{moments}), representing particle number and energy-momentum conservation, must be supplemented by an evolution equation for the dissipative tensor $\Pi^{\mu\nu}$. The traditional way of obtaining the evolution equation is to use (\ref{exp}) in the third equation of (\ref{moments}). In this way the Israel-Stewart equations are obtained. Recently, Denicol et al \cite{den} reobtained these equations directly from the kinetic theory definition of the derivative of $\Pi^{\mu\nu}$ instead of relying on the second moment equation. We note that the form of the equations obtained by these authors is the same as that of Israel-Stewart, but with different transport coefficients which result in better agreement with Boltzmann equation. In any case, a detailed analysis of the transport equation is necessary.

Luzum and Ollitrault \cite{luz} point out that the majority of works in this area assume that $C^{ij}$ is $p$-independent, therefore $C^{ij}=\Pi^{ij}/2\left\langle \omega_p^{4}\right\rangle$. Comparing to eq. (\ref{zeta1sol}) above, this corresponds to the case where $F=T$. A more detailed analysis (\cite{dus,luz}) shows that while this obtains in some cases, such as a $\lambda\phi^4$ theory (\cite{dus,libro}), it is not a good description of a hot gluon plasma. Moreover, the usual analysis also assumes for $\Pi^{ij}$ a gradient expansion

\be
\Pi^{ij}=-\eta\sigma^{ij}+\ldots
\label{gradient}
\te
where $\sigma^{ij}$ is the shear tensor, defined in the rest frame as

\be
\sigma^{ij}=u^{i,j}+u^{j,i}-\frac 23\delta^{ij}u^i_{,i}
\label{shear}
\te
and $\eta\propto T^3$ is the shear viscosity.

On the other hand, \cite{dus,luz} also analyze more general options for $C^{ij}$. Concretely, they analyze cases where $C^{ij}\propto \omega_p^{-\alpha}$, with $\alpha =0$, $1/2$ and $1$. In the EPVM framework they correspond to $F =T\left(\omega_p/T\right)^{-\alpha}$. So $\alpha =0$ corresponds to Marle's kinetic equation, while $\alpha =1$ gives the equation proposed by Anderson and Witting \cite{ander,taka}). 

In summary, we have found a way to associate a simple kinetic equation (which is nevertheless consistent with energy-momentum conservation and the Second Law) to the non-standard nonequilibrium corrections investigated in \cite{dus,luz}. Moreover, our treatment nowhere assumes a gradient expansion such as (\ref{gradient}). We will discuss this point further in next Section.

We believe that the form of $\chi$ we have found here (even at first order), being more flexible than the way Grad's ansatz is usually implemented, may be useful to improve the description of freeze-out in heavy-ion collisions, at least in a phenomenological fashion.

\section{EPVM as a divergence-type theory}
\label{dtt}

In this section we give a brief overview of divergence-type theories \cite{geroch,calz98,liu,marcprd,reula}, and make explicit the connection between DTTs and the EPVM.

\subsection{Divergence type theories}

According to Geroch and Lindblom \cite{geroch}, the hydrodynamical description of a nonequilibrium state requires, besides the particle current $N_\mu$ and the stress-energy tensor $T_{\mu\nu}$, a new third order current $A_{\mu\nu\rho}$. These currents are obtained as derivatives of a generating current $\chi^{\mu}$ with respect to the hydrodynamic variables $\alpha =\mu /T$, $\beta^{\mu}=u^{\mu}/T$ and $\xi^{\mu\nu}$, where $\mu$ is the chemical potential (which vanishes identically for a conformal theory), $T$ is the temperature (see below), $u^{\mu}$ the velocity and the nonequilibrium tensor $\xi^{\mu\nu}$ is symmetric and traceless. For a conformal theory $\xi^{\mu\nu}$ is also transverse $\xi^{\mu\nu}u_{\nu}=0$. The relevant equations for a conformal theory are then

\be
T^{\mu\nu}=\frac{\partial \chi^{\mu}}{\partial\beta_{\nu}} ~~ \textrm{and}
\label{eqno1}
\te

\be
A^{\mu\nu\rho}=\frac{\partial \chi^{\mu}}{\partial\xi_{\nu\rho}} ~~ .
\label{eqno2}
\te
The entropy flux is

\be
S^{\mu}=\chi^{\mu}-\beta_{\nu}T^{\mu\nu}-\xi_{\nu\rho}A^{\mu\nu\rho} ~~ .
\label{eqno3}
\te
Assuming that $T^{\mu\nu}$ is conserved, the entropy production is

\be
S^{\mu}_{,\mu}=-\xi_{\nu\rho}A^{\mu\nu\rho}_{,\mu}
\label{eqno4}
\te
so knowledge of the entropy production gives an equation for $A^{\mu\nu\rho}_{,\mu}$ and thus determines the evolution. 

Because $T^{\mu\nu}$ is symmetric, $\chi^{\mu}$ must derive from a potential

\be
\chi^{\mu}=\frac{\partial \chi}{\partial\beta_{\mu}} ~~ .
\label{eqno5}
\te
The most general conformally invariant potential containing up to quadratic terms in the nonequilibrium tensor and yielding a traceless energy-momentum tensor is

\be
\chi = aT^2+T^{-2}\xi_{\tau\lambda}u^{\tau}u^{\lambda}-cT^{-6}\left[\xi_{\tau\lambda}\xi^{\tau\lambda}+24\xi_{\tau\theta}\xi^{\theta}_{\lambda}u^{\tau}u^{\lambda}+168\left(\xi_{\tau\lambda}u^{\tau}u^{\lambda}\right)^2\right] ~~ .
\label{eqno6}
\te
This form of the potential assumes that $\xi^{\mu\nu}$ has dimensions of $T^4$. We retain terms involving $\xi^{\mu\nu}u_{\nu}$, which are zero ``on shell'' but contribute to the derivatives of the potential. The derivatives are computed according to the rules

\be
\frac{\partial T}{\partial\beta_{\mu}}= T^3\beta^{\mu}=T^2u^{\mu} ~~ ,
\te

\be
\frac{\partial u^{\mu}}{\partial\beta_{\nu}}= T\Delta^{\mu\nu} ~~ \textrm{and}
\te

\be
\Delta^{\mu\nu}=g^{\mu\nu}+u^{\mu}u^{\nu}
\te
whereby we get

\bea
\chi^{\mu} &=&u^{\mu} \left\{2aT^3-2T^{-1}\xi_{\tau\lambda}u^{\tau}u^{\lambda}+6cT^{-5}\left[\xi_{\tau\lambda}\xi^{\tau\lambda}+24\xi_{\tau\theta}\xi^{\theta}_{\lambda}u^{\tau}u^{\lambda}+168\left(\xi_{\tau\lambda}u^{\tau}u^{\lambda}\right)^2\right]\right\}\nn
&+&2\Delta^{\mu\lambda}\left\{T^{-1}\xi_{\tau\lambda}u^{\tau}-24cT^{-5}\left[\xi_{\tau\theta}\xi^{\theta}_{\lambda}u^{\tau}+14\left(\xi_{\tau\theta}u^{\tau}u^{\theta}\right)\xi_{\phi\lambda}u^{\phi}\right]\right\}
\label{genchi}
\tea
and, assuming a transverse $\xi^{\mu\nu}$

\be
T^{\mu\nu} =\rho \left[u^{\nu}u^{\mu} +\frac13 \Delta^{\mu\nu} \right]+2\xi^{\mu\nu}-48cT^{-4}\left[\xi^{\mu}_{\tau}\xi^{\tau\nu}-\frac13 \Delta^{\mu\nu}\xi_{\tau\lambda}\xi^{\tau\lambda}\right]
\label{tdtt}
\te
where

\be
\rho =6aT^4-30cT^{-4}\xi_{\tau\lambda}\xi^{\tau\lambda}
\label{rhodtt}
\te
and

\bea
A^{\mu\tau\lambda}&=&T^{-1}\left(g^{\mu\lambda}u^{\tau}+g^{\mu\tau}u^{\lambda}-\frac12g^{\tau\lambda}u^{\mu}\right)\nn
&-&24cT^{-5}\left(\xi^{\mu\lambda}u^{\tau}+\xi^{\mu\tau}u^{\lambda}-\frac12\xi^{\tau\lambda}u^{\mu}\right) ~~ .
\tea
The entropy flux becomes

\be
S^{\mu} =\beta^{\mu} \left\{8aT^4-36cT^{-4}\xi_{\tau\lambda}\xi^{\tau\lambda}\right\} ~~ .
\label{sflux}
\te
We see from eq. (\ref{rhodtt}) that $T$ is not the temperature as measured by a comoving observer. We have already encountered this situation in subsection \ref{sos}. As in there the solution lies in a temperature shift (cfr. eq. (\ref{tshift})).
This redefinition of the temperature represents a correction of order $\xi^4$ in eq. (\ref{tdtt}) and of order $\xi^3$ in $A^{\mu\nu\delta}$, which we are going to neglect since they correspond to terms that would be obtained from a cubic generating function (see \cite{nosprc}). 

\subsection{To DTTs from EPVMs}

We will now show that the EPVM leads to the DTT discussed before. The idea is to seek a solution for $\xi^{\mu\nu}$ as an expansion in the small parameter $\tau$. Therefore we write (compare to eq. (\ref{expandepvm}))

\be
\xi_{\mu\nu}=\xi^{(1)}_{\mu\nu}+\xi^{(2)}_{\mu\nu} ~.
\label{expanddtt}
\te
Matching the first order correction to $T^{\mu\nu}$ in both theories we get

\be
\xi^{\left(1\right)ij}=\frac{\tau}{15}\left\langle \frac{\omega_p^4}F\right\rangle\lambda^{(0)ij} ~~ .
\label{xi1dtt}
\te
We use this result to compute the first nontrivial correction to the entropy flux and match to eq. (\ref{entroepvm}). This determines the $c$ coefficient

\be
c=\frac{5T^5}{12}\left\langle \frac{\omega_p^4}F\right\rangle^{-2}\left\langle \frac{\omega_p^5}{F^2}\right\rangle ~~ .
\label{cdtt}
\te
Knowing $c$, we can math the full $T^{\mu\nu}$ to get

\be
\xi^{\left(2\right)ij}=\frac12\left[K+\frac 4{45} T\tau^2\left\langle \frac{\omega_p^5}{F^2}\right\rangle\right]\left\{\lambda_{m}^{(0)i}\lambda^{(0)mj}-\frac13\delta^{ij}\lambda^{(0)}_{mn}\lambda^{(0)mn}\right\}
\label{xi2dtt}
\te
where $K$ is defined in eq. (\ref{kconst}).

The only remaining step is to find the equation of motion for $\xi^{\mu\nu}$ by matching the entropy production as given in the DTT to the corresponding result from the EPVM, eq. (\ref{divs1}). This gives (in the rest frame)

\be
\dot\xi^{ij}=-\Gamma\left[\frac{\eta}2\sigma^{ij}+\xi^{ij}\right]-\left[5\frac{\dot T}T+\frac13u^k_{,k}\right]\xi^{ij}+\xi^{i}_k\sigma^{kj}+\xi^{j}_k\sigma^{ki}-\frac 23\delta^{ij}\xi_{kl}\sigma^{kl}
\te
where

\be
\eta =\frac{\tau}{15T}\left\langle \frac{\omega_p^4}F\right\rangle ~~ \textrm{and}
\te

\be
\Gamma =\frac 6{\tau}\left\langle \frac{\omega_p^5}{F^2}\right\rangle^{-1}\left\langle \frac{\omega_p^4}F\right\rangle ~~ .
\te
We recover eq. (\ref{gradient}) when $\tau\to 0$.

\section{Summary}
\label{summ}
Relying on entropy production variational methods and using a linear collision operator that satisfies the $H$ theorem, we have shown how to associate a one particle distribution function to the energy momentum and entropy currents of a conformal fluid, in a way that generalizes Grad's ansatz. The entropy production variational method leads to a nonequilibrium correction to the distribution function which at first order, and for a specific form of the linear collision operator, reproduces Grad's quadratic ansatz. For other choices of the collision operator we obtain a generalization of Grad's ansatz, in which the nonequilibrium distribution function can have a dependence on momentum other than quadratic as indicated by recent developments \cite{dus,luz}. 

Moreover, by equating the entropy production to its extremal value we obtain a hydrodynamics which belongs to the class of divergence-type theories. The kinetic origin of the DTT ensures that the Second Law is satisfied. The derivation provides a kinetic interpretation of the DTT nonequilibrium tensor as a Lagrange multiplier enforcing the stress-energy constraint when extremizing the entropy production. 

Although based on a linear collision term, we believe that our results provide an interesting link between kinetic theory and dissipative divergence-type theories which may prove useful in the study of heavy-ion collisions, for instance, to improve the description of the freeze-out process. It would be interesting to find the dependence of the relaxation time on energy by matching the results of hydrodynamic simulations of heavy-ion collisions to data, in this way constraining the form of the linear collision operator used in the kinetic description of relativistic plasmas.

\begin{acknowledgments}
This work has been supported in part by ANPCyT, CONICET and UBA under project UBACYT X032 (Argentina), and by FAPESP (Brazil).
\end{acknowledgments}


\begin{thebibliography}{999}
\bibitem{is72} W. Israel, {\it The relativistic Boltzmann equation}, in L. O'Raifeartaigh (ed.), {\it General
Relativity: Papers in Honour of J. L. Synge} (Clarendon Press, Great Britain, 1972).
\bibitem{libro} E. Calzetta and B.-L. Hu, {\it Nonequilibrium Quantum Field Theory} (Cambridge University Press, Great Britain, 2008).
\bibitem{first} C. Eckart, Phys. Rev. {\bfseries 58}, 919 (1940); L. Landau, and E. Lifshitz, {\it Fluid Mechanics} (Pergamon Press, Great Britain, 1959).
\bibitem{luz} M. Luzum and J.-Y. Ollitrault, Phys. Rev. C {\bfseries 82}, 014906 (2010).
\bibitem{dus} K. Dusling, G. Moore, and D. Teaney, arXiv:0909.0754 [nucl-th].
\bibitem{mon} A. Monnai and T. Hirano, Phys. Rev. C {\bfseries 80}, 054906 (2009); A. Monnai, and T. Hirano, arXiv:1003.3087 [nucl-th]; G. S. Denicol, T. Kodama, T. Koide, and Ph. Mota, Phys. Rev. C {\bfseries 80}, 064901 (2009).
\bibitem{den} G.S. Denicol, T. Koide, and D.H. Rischke, arXiv:1004.5013v1 [nucl-th].
\bibitem{rom09} P. Romatschke, Int. J. Mod. Phys. E {\bfseries 19}, 1 (2010);  D. T. Son, and A. O. Starinets, Annual Review of Nuclear and Particle Science, {\bfseries 57}, 95 (2007).
\bibitem{israel} W. Israel, Ann. Phys. (NY) {\bfseries 100}, 310 (1976); W. Israel, and J. Stewart, Ann. Phys. (N.Y.) {\bfseries 118}, 341 (1979).
\bibitem{groot} S. R. de Groot, W. A. van Leeuwen and Ch. G. van Weert, {\it Relativistic Kinetic Theory} (North-Holland, Netherlands, 1980). 
\bibitem{stew} John M. Stewart, {\it Non-Equilibrium Relativistic Kinetic Theory}, Lecture Notes in Physics 10 (Springer, Germany, 1971).
\bibitem{gorban} A.N. Gorban and I.V. Karlin, {\it Invariant Manifolds for Physical and Chemical Kinetics} (Springer, Germany, 2005).
\bibitem{grad} H. Grad, Commun. Pure Appl. Math. {\bfseries 2}, 331 (1949).
\bibitem{chap} S. Chapmann and T. G. Cowling, {\it The Mathematical Theory of Non-Uniform Gases} (Cambridge University Press, Great Britain, 1970).
\bibitem{cerc} C. Cercignani, and G. M. Kremer, {\it The relativistic Boltzmann equation: theory and applications} (Birkh$\ddot{a}$user, Germany, 2002).
\bibitem{ktoh} R. Baier, P. Romatschke, and U. A. Wiedemann, Phys. Rev. C {\bfseries 73}, 064903 (2006);  B. Betz, D. Henkel, and D. H. Rischke, J. Phys. G: Nucl. Part. Phys. {\bfseries 36}, 064029 (2009); M. A. York and G. D. Moore, Phys. Rev. D {\bfseries 79}, 054011 (2009); A. Muronga, Phys. Rev. C {\bfseries 76}, 014910 (2007); X. Chen, H. Rao, and E. A.  Spiegel, Phys. Rev. E {\bfseries 64}, 046308 (2001); E. A. Calzetta, B.-L. Hu, and S. A. Ramsey, Phys. Rev. D {\bfseries 61}, 125013 (2000). 
\bibitem{mep} L. M. Martyushev, and V. D. Seleznev, Phys. Rep. {\bfseries 426}, 1 (2006).
\bibitem{chris} T. Christen, EPL {\bfseries 89}, 57007 (2010).
\bibitem{nos} J. Peralta-Ramos and E. Calzetta, Phys. Rev. D {\bfseries 80}, 126002 (2009). 
\bibitem{nosprc} J. Peralta-Ramos and E. Calzetta, arXiv:1003.1091 [hep-ph].
\bibitem{geroch} R. Geroch and L. Lindblom, Phys. Rev. D {\bfseries 41}, 1855 (1990).
\bibitem{jou} D. Jou and D. Pavon, Phys. Rev. A {\bfseries 44}, 6496 (1991); I. Bouras, E. Molnar, H. Niemi, Z. Xu, A. El, O. Fochler, C. Greiner and D.H. Rischke, arXiv:1006.0387 [hep-ph]; ibid, Nucl. Phys. A {\bfseries 830}, 741 (2009).
\bibitem{brss} R. Baier, P. Romatschke, D. T. Son, A. O. Starinets, and M. A. Stephanov, J. High Energy Phys. {\bfseries 04}, 100 (2008); S. Bhattacharyya, V. E. Hubeny, S. Minwalla, and M. Rangamani, J. High Energy Phys. {\bfseries 02}, 45 (2008); M. Natsuume and T. Okamura, Phys. Rev. D {\bfseries 77}, 066014 (2008); Erratum-ibid. D {\bfseries 78}, 089902 (2008).
\bibitem{marle} C. Marle, C. R. Acad. Sc. Paris {\bfseries 260}, 6539 (1965).
\bibitem{ander} J. L. Anderson, and H. R. Witting, Physica {\bfseries 74}, 466, (1974).
\bibitem{taka} M. Takamoto, and S.-I. Inutsuka, arXiv:1006.2663 [astro-ph.HE].
\bibitem{calz98} E. Calzetta, Class. Quant. Grav. {\bfseries 15}, 653 (1998).
\bibitem{liu} I.-S. Liu, I. Muller, and T. Ruggeri, Ann. Phys. (N.Y.) {\bfseries 169}, 191 (1986); T. Ruggeri, in Lecture Notes in Mathematics Vol. 1385, Eds. A. Anile and Y. Choquet-Bruhat (Springer-Verlag, Germany, 1989).
\bibitem{marcprd} E. Calzetta, and M. Thibeault, Phys. Rev. D {\bfseries  63}, 103507 (2001).
\bibitem{reula} O. A. Reula, and G. B. Nagy, J. Phys. A {\bfseries 30}, 1695 (1997). 


\end{thebibliography}
\end{document}